\documentclass[]{spie}  

\usepackage{amsmath,amsfonts,amssymb}
\usepackage{graphicx,color}
\usepackage{lineno}
\usepackage{aas_macros}

\title{CAMELOT: design and performance verification of the detector concept and localization capability}

\author[a]{Masanori Ohno}
\author[b,c,a]{Norbert Werner}
\author[e]{Andr{\' a}s P{\' a}l}
\author[b,d]{Jakub {\v R}{\' i}pa}
\author[g]{Gab{\' o}r Galg{\' o}czi}
\author[f]{Norbert Tarcai}
\author[f]{Zsolt V{\' a}rhegyi}
\author[a]{Yasushi Fukazawa}
\author[a]{Tsunefumi Mizuno}
\author[a]{Hiromitsu Takahashi}
\author[a]{Koji Tanaka}
\author[a]{Nagomi Uchida}
\author[a]{Kento Torigoe}
\author[h]{Kazuhiro Nakazawa}
\author[i]{Teruaki Enoto}
\author[j]{Hirokazu Odaka}
\author[k]{Yuto Ichinohe}
\author[g,l]{Zsolt Frei}
\author[e]{L{\' a}sz{\' o} Kiss}

\affil[a]{School of Science, Hiroshima University, 1-3-1 Kagamiyama, Higashi-Hiroshima, Japan }
\affil[b]{MTA-E\"ot\"vos University Lend\"ulet Hot Universe Research Group, P\'azm\'any P\'eter s\'et\'any 1/A, Budapest, 1117, Hungary}
\affil[c]{Department of Theoretical Physics and Astrophysics, Faculty of Science, Masaryk University, Kotl\'a\v{r}sk\'a 2, Brno, 611 37, Czech Republic }
\affil[d]{Charles University, Faculty of Mathematics and Physics, Astronomical Institute, V Hole\v{s}ovi\v{c}k\'ach 2, 180 00 Prague 8, Czech Republic }
\affil[e]{Konkoly Observatory of the Hungarian Academy of Sciences, Konkoly-Thege ut 15-17, Budapest, 1121, Hungary}
\affil[f]{C3S Electronics Development LLC., K{\"o}nyves K\'alm\'an krt. 12-14., Budapest, 1097, Hungary}
\affil[g]{Institute of Physics, E\"otv\"os University, P\'azm\'any P\'eter s\'et\'any 1/A, Budapest, 1117, Hungary}
\affil[h]{Department of Physics, Nagoya University, Furo-cho, Chikusa-ku, Nagoya, Aichi, Japan}
\affil[i]{The Hakubi Center for Advanced Research and Department of Astronomy, Kyoto University, Kyoto 606-8302, Japan}
\affil[j]{Department of Physics, University of Tokyo, 7-3-1 Hongo, Bunkyo, Tokyo 113-0033, Japan}
\affil[k]{Department of Physics, Rikkyo University, Nishi Ikebukuro 3-34-1, Toshimaku, Tokyo 171-8501, Japan}
\affil[l]{MTA-ELTE Astrophysics Research Group, P\'azm\'any P\'eter s\'et\'any 1/A, Budapest, 1117, Hungary}

\authorinfo{Further author information: (Send correspondence to M. Ohno)\\M. Ohno: E-mail: ohno@astro.hiroshima-u.ac.jp}

\begin{document} 
\maketitle

\begin{abstract}
A fleet of nanosatellites using precise timing synchronization provided by the Global Positioning System is a new concept for monitoring the gamma-ray sky that can achieve both all-sky coverage and good localization accuracy. We are proposing this new concept for the mission CubeSats Applied for MEasuring and LOcalising Transients (CAMELOT). The differences in photon arrival times at each satellite are to be used for source localization. Detectors with good photon statistics and the development of a localization algorithm capable of handling a large number of satellites are both essential for this mission. Large, thin CsI scintillator plates are the current candidates for the detectors because of their high light yields. It is challenging to maximize the light-collection efficiency and to understand the position dependence of such thin plates. We have found a multi-channel readout that uses the coincidence technique to be very effective in increasing the light output while keeping a similar noise level to that of a single channel readout. Based on such a detector design, we have developed a localization algorithm for this mission and have found that we can achieve a localization accuracy better than 20 arc minutes and a rate of about 10 short gamma-ray bursts per year.

\end{abstract}


\section{INTRODUCTION}
\label{sec:intro}  

There is no doubt that the era of "multi-messenger astronomy" has now begun. Five gravitational-wave sources originating from blackhole-blackhole mergers have been detected since the first detection in September 2014 by the Laser Interferometer Gravitational-Wave Observatory (LIGO) in the United States\cite{PhysRevLett.116.061102,PhysRevLett.116.241103,2017PhRvL.118v1101A,PhysRevLett.119.141101,2041-8205-851-2-L35}. In addition, high-energy cosmic-neutrino observations with quick-detection alerts have been started recently by the IceCube Neutrino Observatory. Thanks to the progress of such observational techniques, in 2017, we were finally able to detect the electromagnetic counterparts of both a gravitational-wave and a cosmic-neutrino source\cite{2017ATel10791....1T}. This was truly the dawn of the "multi-messenger astronomy" era.
The detection of the electromagnetic counterpart to the gravitational-wave source GW170817, which resulted from a  double neutron-star merger, really opened our eyes to the potentials of multi-messenger astronomy. 
A gigantic campaign of the follow-up observations at many different electromagnetic wavelengths was successfully conducted, and  it confirmed that the neutron-star merger was the site of the production of heavier, "r-process", elements, as the behavior of the optical light curve and the spectrum were both consistent with  predictions for a "kilo-nova"\cite{2017PASJ...69..101U}. A long monitoring campaign of the X-ray emission from this source is still underway, and an indication has been found  of the decay of the X-ray emission, which may yield important information about the detailed structure of the jet\cite{2017Natur.551...71T,2018ATel11619....1T}, which is the narrowly collimated, ultra-relativistic particle flow produced by the neutron-star merger. 
While this electromagnetic follow-up observation campaign has been very successful, several challenges remain for future follow-up observations. The first problem is to investigate direct evidence for the origin of short-duration gamma-ray bursts (SGRBs). Gamma-ray Bursts (GRBs) have been well-known for many years as the most energetic explosions in the universe, but our understanding of their nature, for instance, their progenitors, emission mechanisms, and origins of their ultra-relativistic jets, is still incomplete. A strong candidate for the progenitor of SGRBs is thought to be the on-axis jet emission produced by a double neutron-star merger, which is also a source of gravitational waves. Thus, the simultaneous detection of gravitational waves and gamma-ray emission from a SGRB would provide direct evidence for the nature of the progenitor. Indeed, gamma-ray emission has been also detected from GW170817, but curiously, the gamma-ray luminosity was much lower than that expected for a nominal SGRB. A simple way to address this problem is  to continue gamma-ray observations with a large field of view. We have to maintain a continuous all-sky coverage in order not to miss a chance for a simultaneous gravitational-wave detection. 

Another important issue is quick and efficient follow-up observations. The localization uncertainty of the gravitational-wave detectors is several tens of square degrees, which is much larger than the typical field of view of an optical telescope and, in general, it is impossible to start deep, optical follow-up observations before quick localization information has been provided from other wavelengths. For GW170817, optical follow-up observations started more than  half a day after the detection of the gravitational waves\cite{2017Sci...358.1556C}. Therefore, we lost valuable electromagnetic information from the early phase, which can be important for nucleosynthesis modeling\cite{2017PASJ...69..102T}. 

Considering these challenges for the future electromagnetic follow-up observations of gravitational-wave sources, we propose a new concept for GRB detection that can provide both all-sky coverage and a precise localization to within one square degree that we have named  {\it CubeSats Applied for MEasuring and LOcalising Transients (CAMELOT)}.  The details of this mission concept are described in Werner et al. in this volume\cite{Werner+18SPIE}. A localization accuracy of one square degree is sufficient  for future optical telescopes, which are expected to have a field of view of a similar size. However, it is challenging to obtain both all-sky coverage and precise localization. Our concept is to utilize a fleet of multiple nanosatellites, with a precise timing synchronization provided by Global Positioning System (GPS), as described by P{\' a}l et al. in this volume\cite{Andras+18SPIE}. Once we obtain a sufficient timing synchronization, we can measure the differences in the arrival times of photons from gravitational-wave sources even if the fleet of nanosatellites is deployed in low Earth orbit if we have good enough photon statistics. This will enable us to achieve a localization accuracy of several tens of arc minutes to degrees. In this paper, we will focus on describing the design and performance of our gamma-ray detectors, and on introducing  a framework for estimating the localization accuracy of our mission concept.

\section{DETECTOR DESIGN}

\begin{figure}[htbp]
 \begin{center}
 \rotatebox{-0}{\resizebox{10cm}{!}{\includegraphics{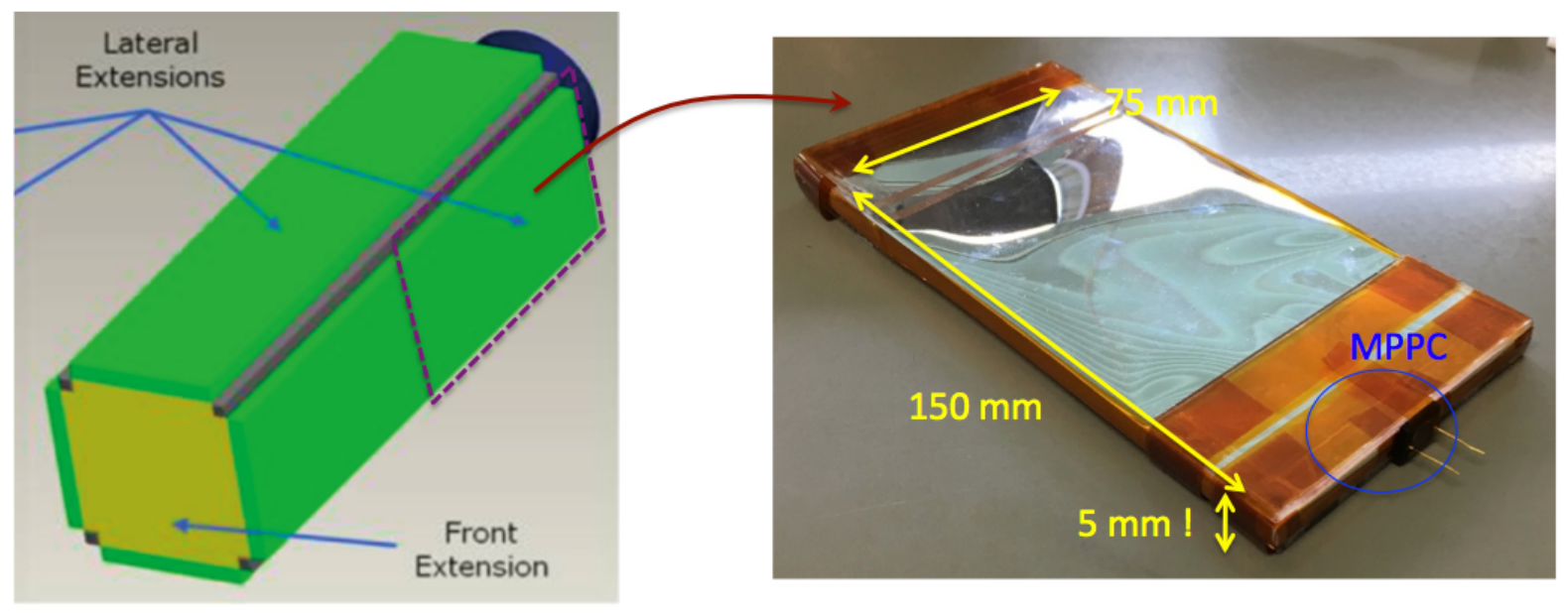}}}
\end{center}
\caption{(Left) Schematic of the detector configuration. (Right) The actual size and material of the CsI detector plate.}
\label{fig:det1}
\end{figure}

As noted in the previous section, high photon statistics is essential for good
localization accuracy. For this reason, the main drivers for our detector design are to maximize the geometrical area and to obtain as low an energy threshold 
as possible. We plan to employ a 3U Cubesat platform, and the lateral extensions are good candidates for the location of the detectors, as shown in the figure \ref{fig:det1}  left. 
There are two separate spaces on one lateral side where we can place our detector. The available space per detector is about 150 $\times$ 83 $\times$ 9 mm$^3$.
Four detectors will be installed on two perpendicular sides. We plan to employ a CsI(TI) scintillator as the gamma-ray detector,  because its high light yield will enable us to achieve a lower energy threshold. The thickness of each detector will be about 5 mm, considering the characteristics of the other support structures, which
indicates that we require a very compact readout device for the scintillation photons. 
Multi-Pixel Photon Counters (MPPC) are suitable because of their compact readout area, tolerance of the readout noise, and low operational voltage. Therefore, the basic concept of our current detector design is to employ four large, thin,
plate-shaped CsI(TI) scintillators readout by the MPPCs (figure\ref{fig:det1} right). It is known that the gain of the MPPCs exhibits a strong dependence on temperature.
We plan to monitor characteristic emission lines such as 511 keV due to activated materials and correct the gain based on on-ground analysis, and we hope to be able to adjust the high-voltage actively in-orbit.
It is challenging to use such large, thin scintillators with small readout areas because the generated scintillation photons follow
very complicated paths, undergoing much scattering and absorption before they reach the MPPC readout area.
This may cause degradation and non-uniformity in the yield of the number of readout scintillation 
photons, depending on the position of the incident gamma-rays. Therefore, a careful performance test by using the actual configuration, including the size of the detector and the number and locations of the readout 
devices, is very important for our mission design. In the following sections, we describe the details and current results from the detector-performance tests.

\subsection{Detector configuration for the performance test}

\begin{figure}[htbp]
 \begin{center}
  \rotatebox{-0}{\resizebox{12cm}{!}{\includegraphics{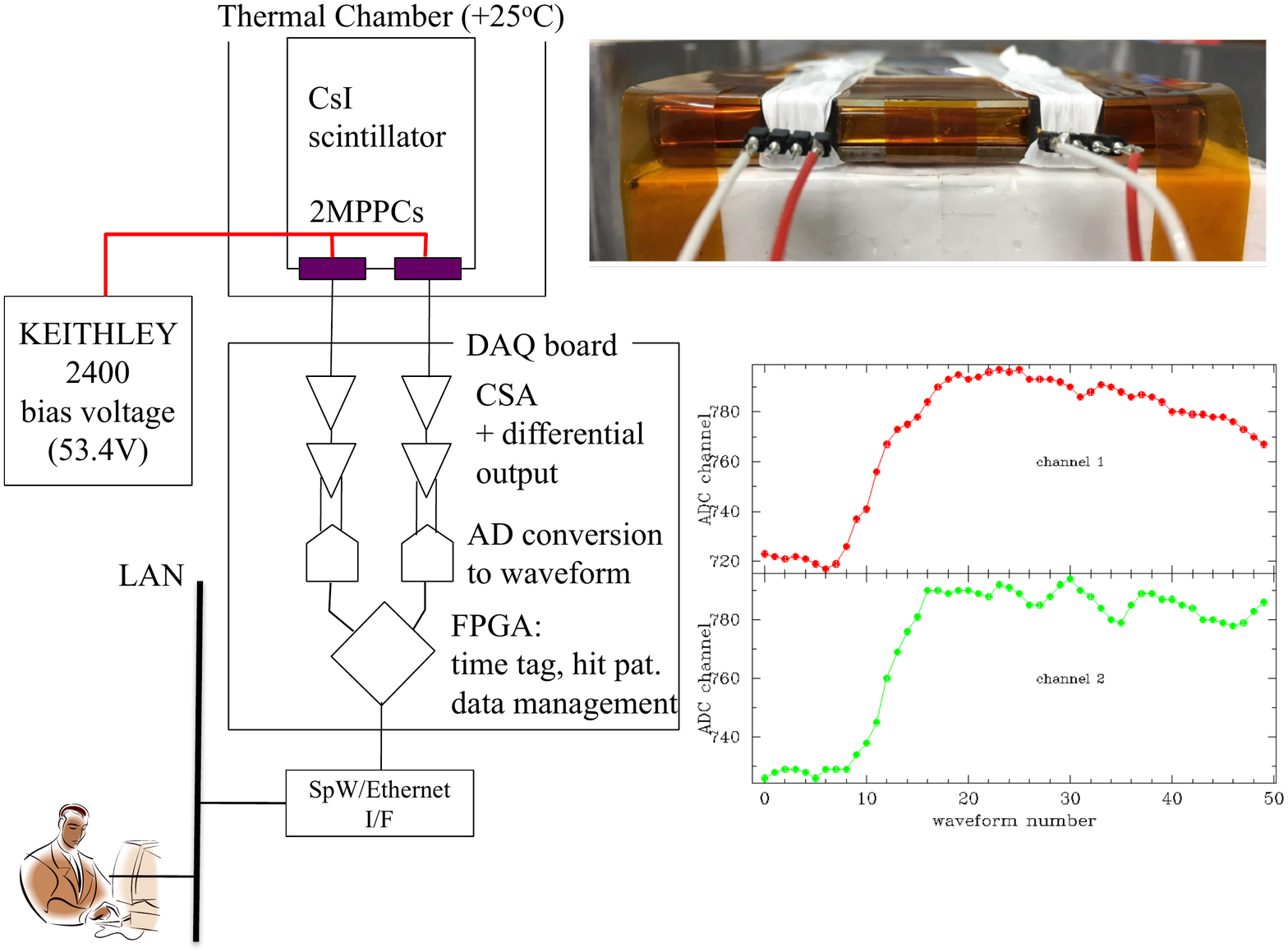}}}
\end{center}
\caption{The MPPC configuration for two-channel readout, together with a functional block diagram.  An example of the obtained waveform data for a coincidence between the two MPPCs is also shown.}
\label{fig:2chan}
\end{figure}

For this performance test, we employed a scintillator with dimensions of 150 $\times$ 75 $\times$ 5 mm$^3$,
read out by two MPPCs (Hamamatsu Photonics S13360-6050CS). These two MPPCs were installed on the same side of the detector, keeping the same separation between each MPPC and the edge of the detector, as shown in the top panel of Figure \ref{fig:2chan}. The operational voltage and temperature were controlled to be
53.4 V and 25 $^\circ$C, respectively. The CsI scintillator was enclosed by 
65$\mu$m of Enhanced Specular Reflector so that the optical scintillation photons do not escape from
the scintillator. Herein, we apply a coincidence readout technique to suppress the increase in dark current that
corresponds to the number of MPPCs. 
In order to determine coincidence, we recorded not only the waveforms from each MPPC but also the trigger time for
each event, and  events for which the trigger times matche to within 5 $\mu$s are regarded as the coincident events. 
The data flow and an example of the waveforms from a coincidence event are shown in the bottom panel of Figure \ref{fig:2chan}.
For this test, we used gamma-rays from an $^{241}$Am radioisotope positioned about 14 mm from the
surface of the scintillator. In order to measure the non-uniformity of the readout scintillation photons,
we placed a collimator made of 1-mm thick lead, which contains 9 pinholes, each 1 mm in diameter. 
 A detailed illustration of this setup can be found in Torigoe et al. (2018)\cite{Torigoe+18NIMA}.

\subsection{Result of the performance evaluation}

In this performance-evaluation test, we have checked the following items: 1) spectral performance, including
energy resolution and energy threshold,  using a two-channel coincidence readout; 2) non-uniformity of the output of the scintillation photons, depending on the position of the incident gamma-rays.

\subsubsection{Spectral performance}
\label{sec:spec}

Figure \ref{fig:exp_Am_all} shows the spectrum obtained by irradiating the detector with an $^{241}$Am radioisotope without
the lead collimator. The peak structure clearly shows the  photo-absorption of the 59.5 keV gamma-rays for both the single channel  readout and the two-channel coincidence readout. Note that even if we use two MPPC devices, the low-energy threshold
does not changed significantly. We achieved a threshold of about 10 keV, while the energy resolution improved from
35.1$\pm$0.7\% to 29.2$\pm$0.4\%, as shown in Torigoe et al. (2018), thanks to the coincidence readout technique.

\subsubsection{Non-uniformity of the scintillation-photon output}

\begin{figure}[htbp]
\begin{center}
\rotatebox{-0}{\resizebox{10cm}{!}{\includegraphics{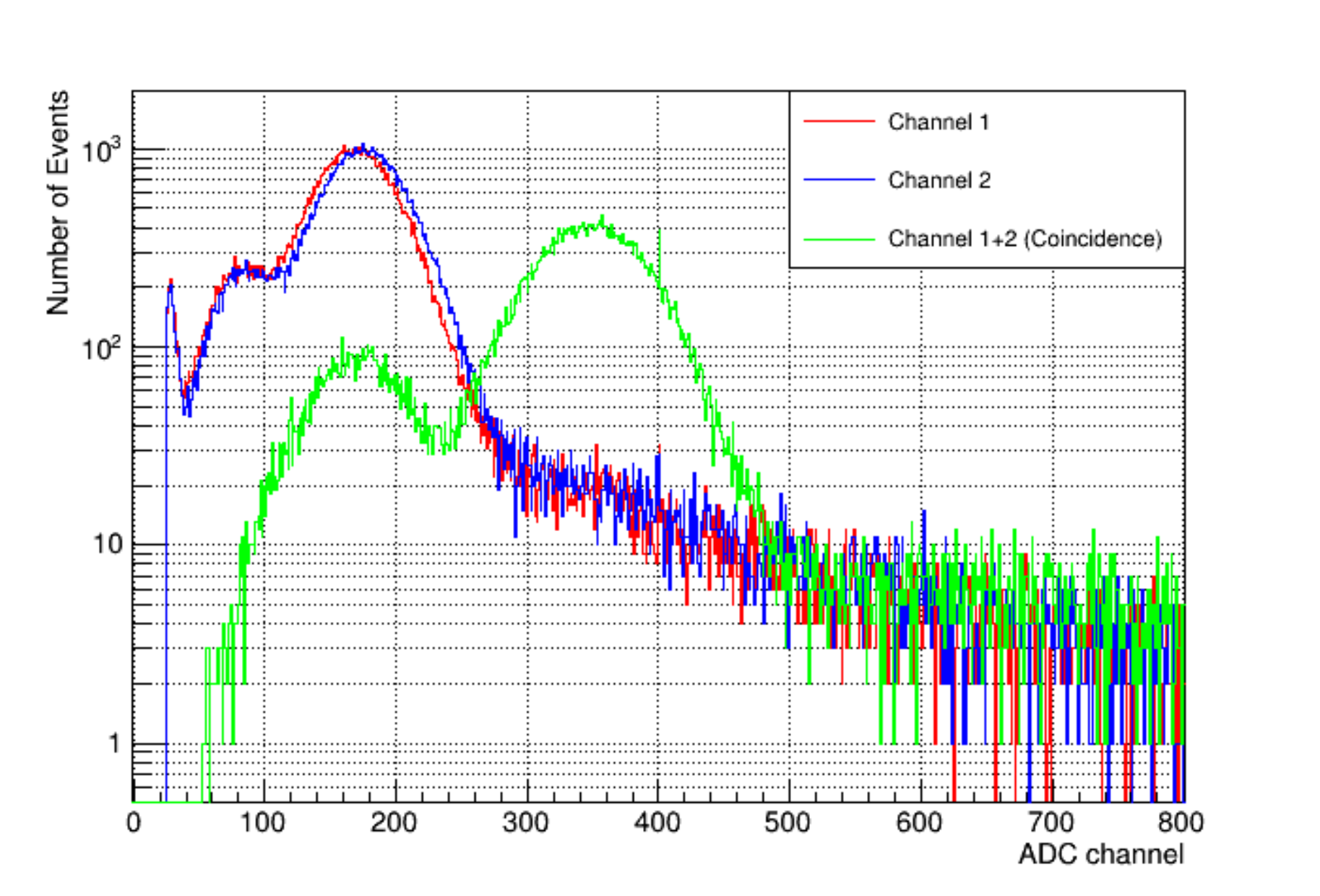}}}
\caption{An example of our detector performance. The spectrum was obtained by irradiating the detector with an $^{241}$Am radioisotope without the collimator. Blue and red data shows the spectrum from each MPPC using a simple single-channel readout, and the result from a two-channel coincidence readout is shown by the green data.}
\label{fig:exp_Am_all}
\end{center}
\end{figure}

\begin{figure}[htbp]
\begin{center}
 \rotatebox{-90}{\resizebox{10cm}{!}{\includegraphics{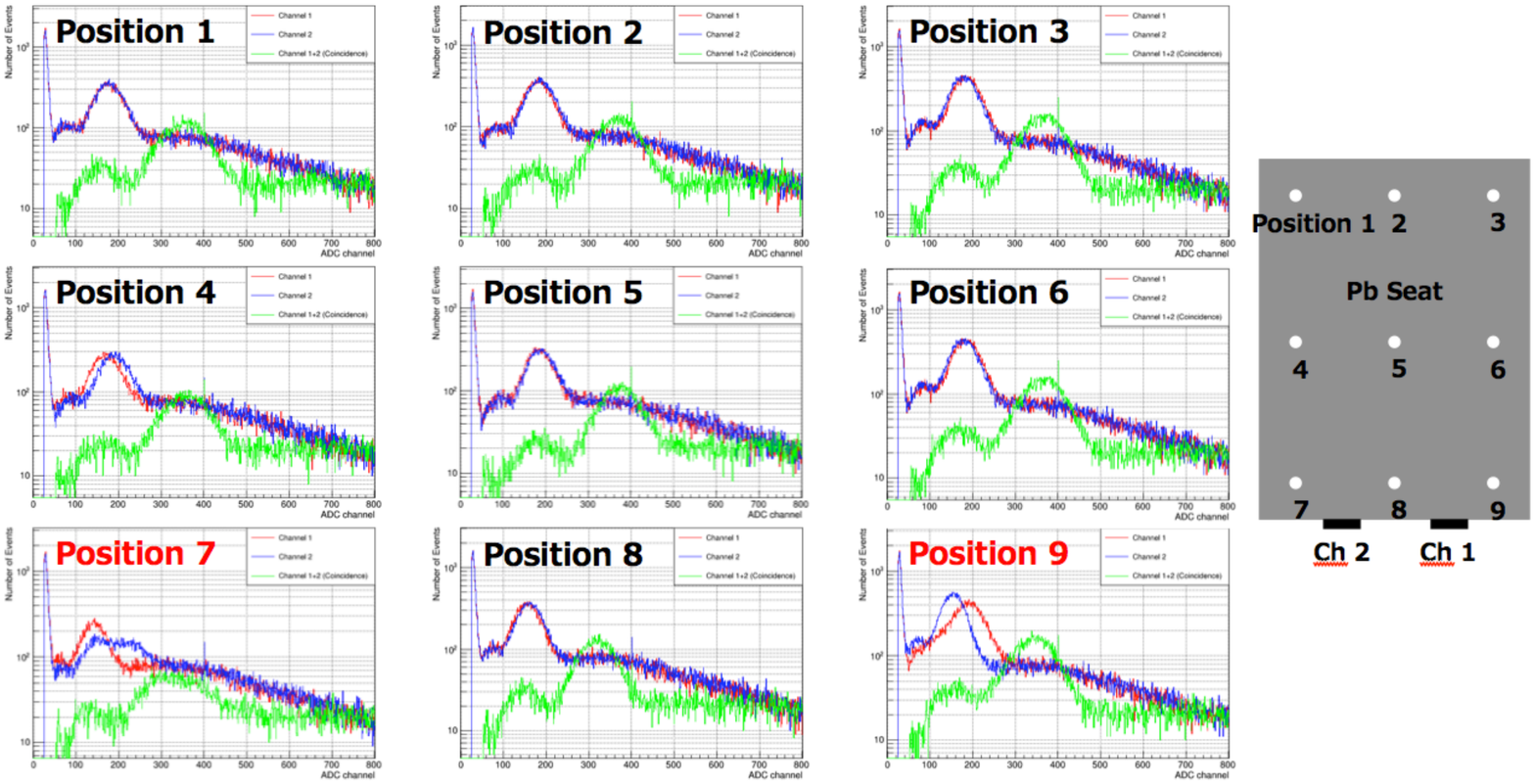}}}
\caption{Same as Figure \ref{fig:exp_Am_all} except that the gamma-rays are passed through  various pinhole positions. The positions of the pinholes are shown in the right-hand side of this figure (modified from the Figure 1 of Torigoe et al. 2018)}
\label{fig:exp_Am_posdep}
\end{center}
\end{figure}

Figure \ref{fig:exp_Am_posdep} is the same as Figure \ref{fig:exp_Am_all} except that we have collimated the  input gamma-rays by using a lead collimator with 9 pinholes. The positions of the pinholes are described in Torigoe et al. (2018) and are shown in the Figure \ref{fig:exp_Am_posdep}.
From the measurement of the non-uniformity of the readout of the scintillation photons, we found that most of the positions do not show a 
strong change for either of the two MPPCs; only the positions closest to the MPPCs show distortions in the spectrum.
This is probably due to the fact that scintillation photons created close to a MPPC readout can reach to the readout area efficiently, compared with other positions, so the resulting light output is significantly different for those positions.
Contamination from the other peak at around 20 keV is produced by the shift of the photo-absorption events of 59.5 keV due to  the escape of fluorescent X-rays of elemental Cs or I.
This indicates that our detectors have very strong non-uniformities only in limited areas  close to the MPPC readout positions.
More detailed investigations of these non-uniformities close to the positions of each MPPC readout, together with further investigations of the number of readouts and the positions of the MPPC devices, should be considered in further experiments
to provide a detailed understanding of the cause of this non-uniformity and to reduce this trend.

\subsection{Angular response of the effective area}

\begin{figure}[htbp]
\begin{center}
\begin{minipage}{8cm}
 \rotatebox{-0}{\resizebox{8cm}{!}{\includegraphics{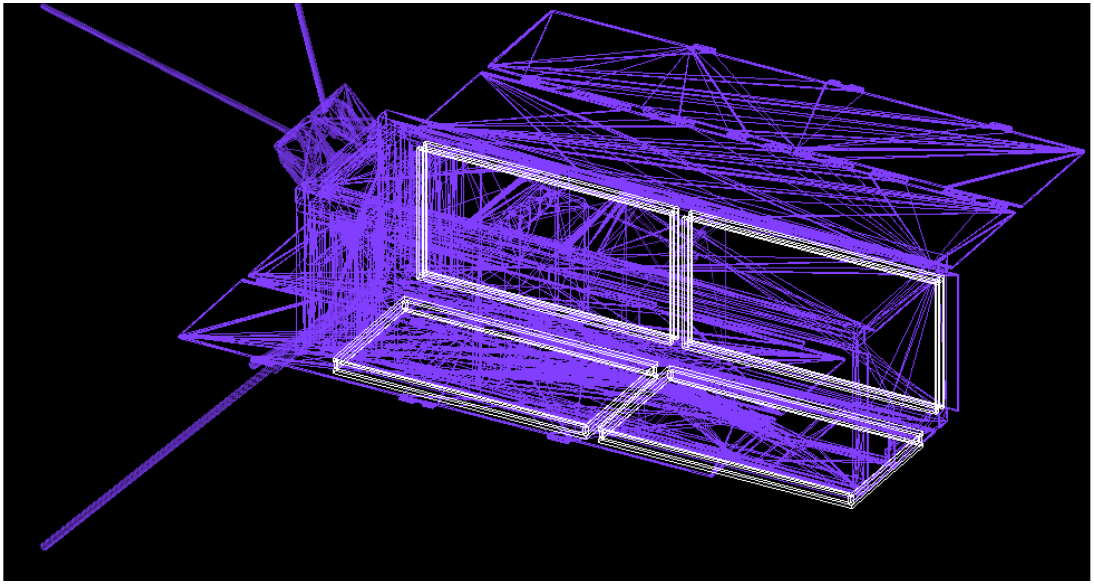}}}
\end{minipage}
\begin{minipage}{8cm}
 \rotatebox{-0}{\resizebox{8cm}{!}{\includegraphics{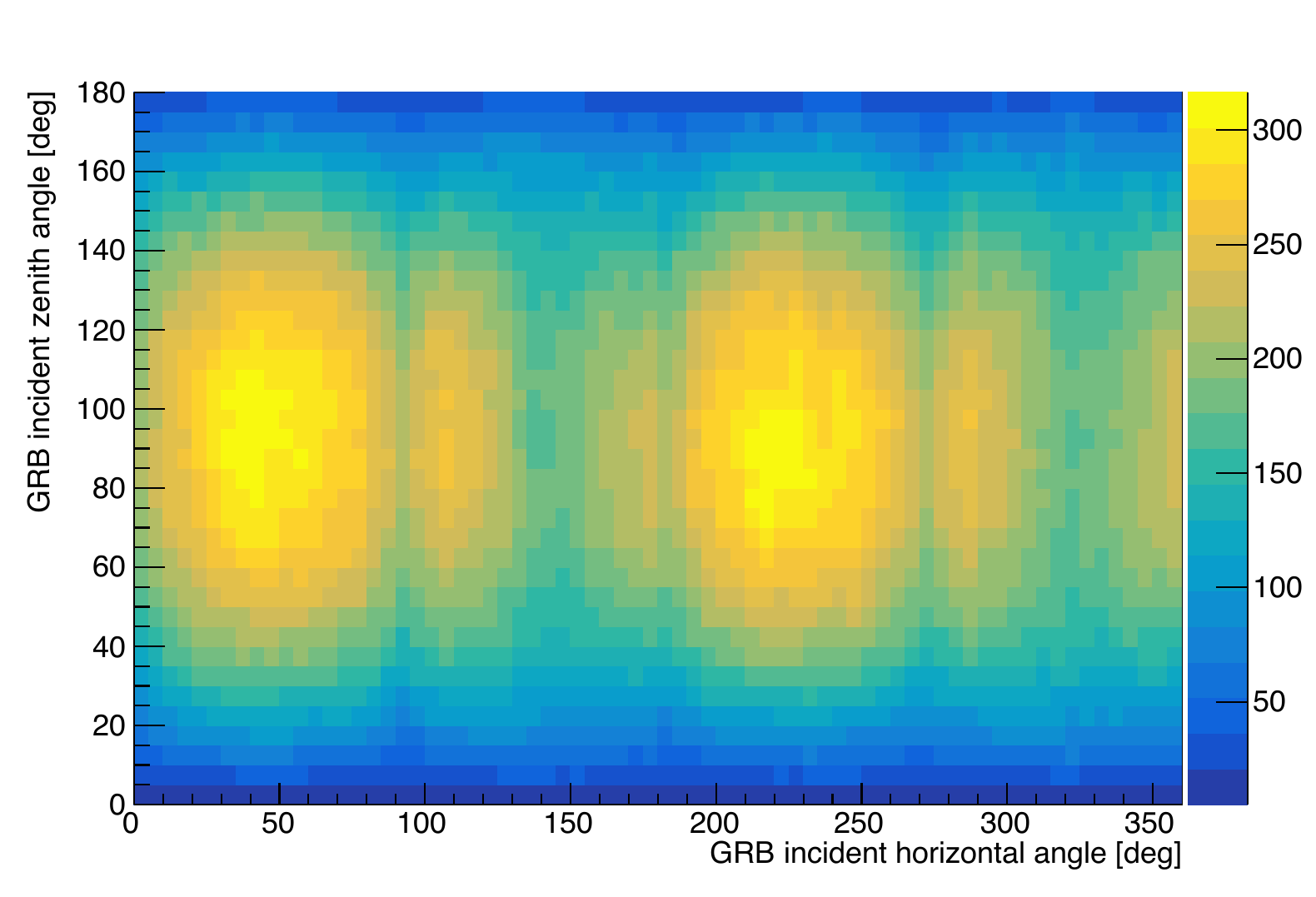}}}
\end{minipage}
\caption{(Left) The mass model of our detector reproduced in the Geant4 simulation. Note that the satellite structure is used for the visualization only and that the current simulation does not
take into consideration the scattering by this structure. (Right) The result of the calculation of the effective area of our detector, as a two-dimensional map plotted as a function of the incident GRB horizontal and
zenith angles relative to the detector coordinates. The color chart shows the effective area in the unit of cm$^2$.}
\label{fig:g4}
\end{center}
\end{figure}

\begin{figure}[htbp]
\begin{center}
 \rotatebox{-0}{\resizebox{10cm}{!}{\includegraphics{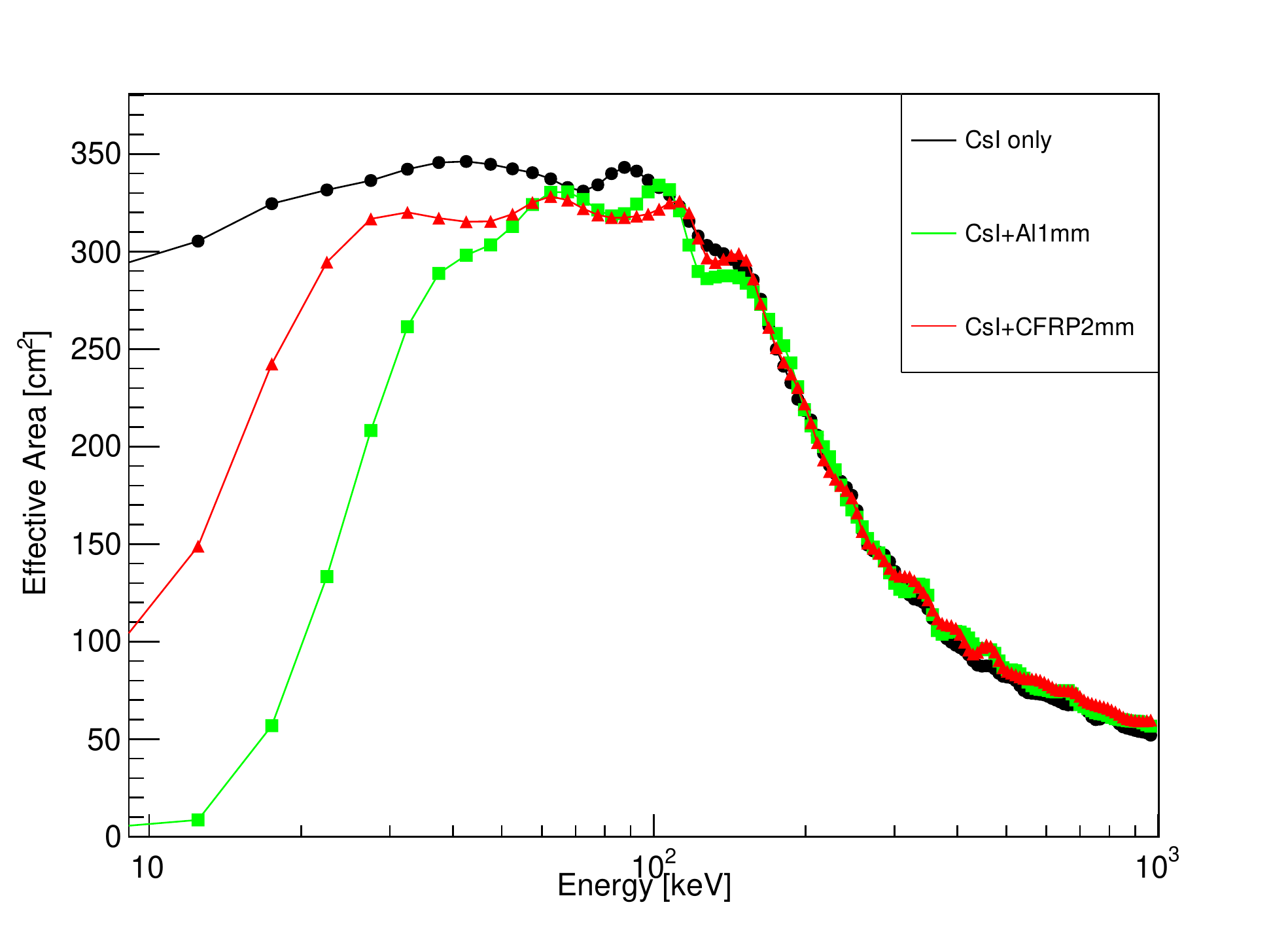}}}
\caption{The energy dependence of the effective area of our detector for the case of the best incident angle. The black circles, red triangles, and green squares show the result for the CsI scintillator only, the CsI scintillator enclosed by a 1-mm thickness of aluminum, and enclosed by a 2-mm thickness of CFRP, respectively.}
\label{fig:effarea_vs_energy}
\end{center}
\end{figure}

As  mentioned in the previous sections, one of the most important parameters of our mission is the effective photon collecting area.
The effective area changes with respect to the direction of the incoming photon. We have to take into consideration such variations when 
we evaluate the localization accuracy of our mission. 
We must therefore evaluate the effective area of our detector for any gamma-ray incident angle. 
For this purpose, we developed a Monte Carlo simulator based on Geant4 version 10.4 that includes 
the actual detector parameters, such as the energy resolution and low-energy threshold of our detector, which we already evaluated, as described in \S\ref{sec:spec}. Figure \ref{fig:g4} (left) shows an
example of our detector simulation. We constructed the detector mass model and we assume an input of 10$^5$ photons with a pseudo-GRB spectrum, which 
has a broken-power-law shape with a low-energy photon index of $-$1.5, a high-energy photon index of $-$2.5 and a break energy of 300 keV, very
similar to a typical GRB photon spectrum. 
Then, we integrated the detected photon numbers over the 10 to 1000 keV energy band and calculated the
effective area. 
We followed these procedures for all GRB incident-angle patterns, in 5-degree steps (2592 patterns), for given detector coordinates. Figure \ref{fig:g4} (right)
shows the effective area map of our detector plotted as a function of its horizontal and zenith angles.
We found that our detector achieves an effective area larger than 300 cm$^2$ for the optimal incident angles, which is comparable to {\it Fermi}-GBM. This effective area map can be used as an input for estimating the expected number of detected photons for the subsequent localization analysis. Figure \ref{fig:effarea_vs_energy} shows the energy dependence of the effective area for the best-incident-angle case. We also checked the effect of the absorption by the support-enclosure material for the CsI scintillator, and we found that to retain a comparable effective area down to 10 keV, it is important to choose the material for the support structure carefully. A 2-mm thickness of carbon-fiber-reinforced polymer (CFRP) is currently the best candidate.

\section{LOCALIZATION ANALYSIS}
 The localization accuracy of a gravitational-wave detector is more than tens of hundreds of  square degrees. Even
 for the best case of GW170817, it was larger than 30 square degrees, and more than 50 host-galaxy candidates are included in such a large error region. The localization information provided by the 
 gamma-ray observations was useless in constraining this error region because of the comparable coarseness of the error region in the gamma-ray energy band for this case. 
 If more accurate localizations, sufficient to constrain the number of host-galaxy candidates to 1-2, can be obtained from gamma-ray observations, faster and more efficient follow-up observations at other 
 wavelengths would enable observations of the early phases of the decay of a kilo-nova, and even of the early phases of the afterglows of the SGRBs, providing strong
  constraints on their origins.
  
In this section, we demonstrate the capability for localizing  GRBs with the {\it CAMELOT} mission by describing the basic concept of
the simulation study and the expected contribution to the future follow-up observations of the gravitational-wave sources. 

\subsection{Analysis strategy}
\begin{figure}[htbp]
\begin{center}
 \rotatebox{-0}{\resizebox{15cm}{!}{\includegraphics{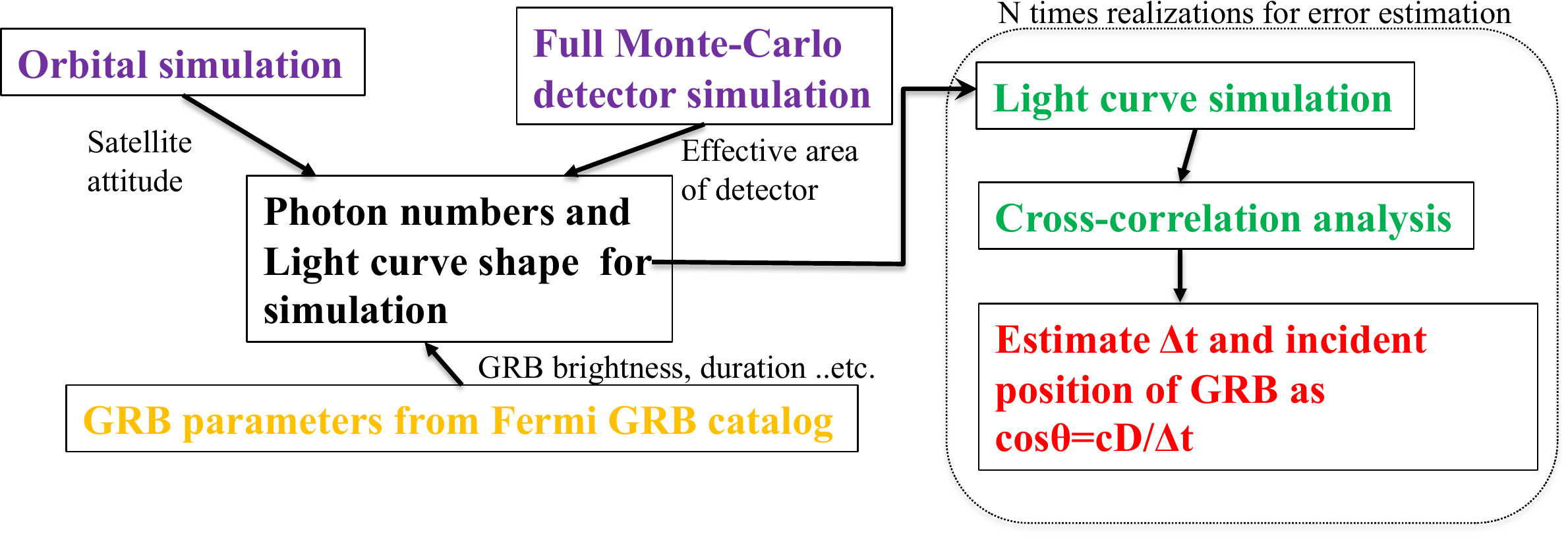}}}
\caption{Block diagram of our framework for the simulated localization analysis}
\label{fig:loc_block}
\end{center}
\end{figure}

The final output from this feasibility study is the expected difference in photon arrival times at each satellite because our basic concept for the localization is triangulation. The triangulation method constrains the incoming photon direction by the simple triangulation formula, cos$\theta$ = c$\delta t/D$, where $c$ is the speed of light, $\delta t$ is the difference in the photon arrival times and $D$ is the distance between a given pair of satellites. This enables us to constrain the direction of the incoming photons along an annulus defined by the angle $\theta$, and the size of error for all parameters in this formula gives the finite width of this annulus. By combining the annuli for some number of satellite combinations,  we can constrain the direction of the incident photon. In this study, we take into account only the statistical error related to $\delta t$. The parameter $\delta t$  can be determined by cross-correlation analysis of the observed light curves. We have, therefore, simulated the observed light curves expected for each satellite with given orbital parameters. The accuracy of the $\delta t$ determination depends on the photon statistics of the light curve. Thus, we have to calculate the effective area of the detector for various GRB incident angles. Finally, we have performed a systematic analysis for actual GRB parameters taken from the  3rd {\it Fermi}-GBM GRB catalog. The concept of this simulation strategy is summarized in Figure \ref{fig:loc_block}.

\subsection{orbital simulation}

To calculate the expected time delays of the photon arrivals and the expected numbers of observed photons, we need to calculate the satellite positions and attitudes in a given time window.
For this calculation, we utilized the results from the orbit simulations described in Werner et al. in this volume. For this localization simulation, we used the parameters for a constellation of 9 satellites on orbits with an inclination of 53 degrees. We also calculated the incident photon direction with respect to the detector coordinates for a given celestial position of the simulated GRB by using
simulated attitude parameters. We define the visibility in terms of the angular distance between the zenith and the direction of the GRB. For the case of this angle distance is larger than 105 degrees, we consider that this satellite cannot observe this GRB due to the occultation by the Earth. 

\subsection{Light curve simulation}

\begin{figure}[htbp]
\begin{center}
 \rotatebox{-0}{\resizebox{15cm}{!}{\includegraphics{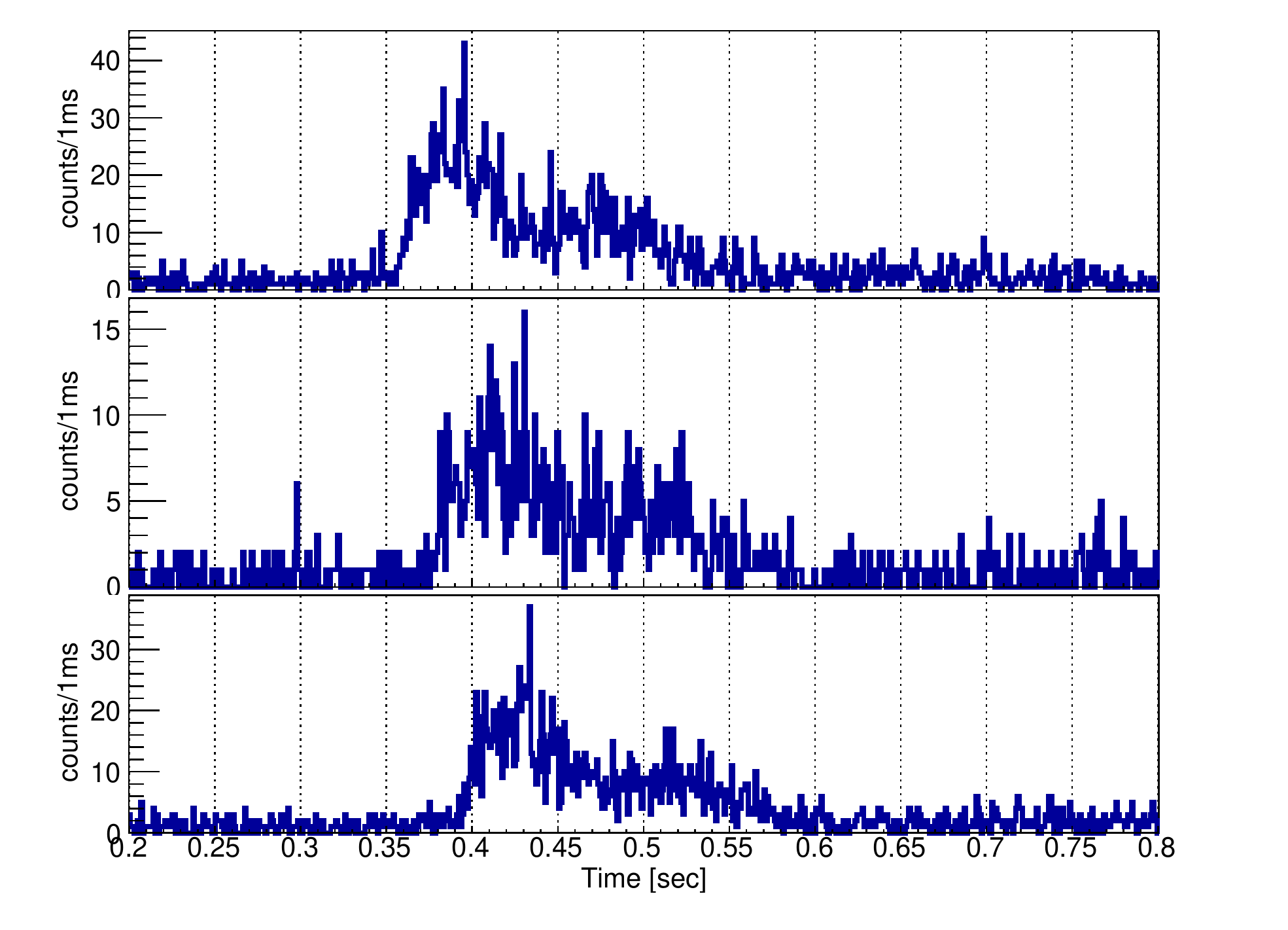}}}
\caption{Examples of simulated light curves for a bright GRB090227772. Three panels show the light curves observed by different satellites.}
\label{fig:lc_sample}
\end{center}
\end{figure}

By using the orbital-simulation parameters, we can determine the GRB incident angles in detector coordinates and the expected photon arrival times for each satellite from a given GRB location and time. The total number of photons that we expect to observe can be determined from an actual light curve observed with the {\it Fermi}-GBM by scaling the ratio of the effective area of our detector from that of the {\it Fermi-GBM}. We can then reconstruct the simulated light curves, assuming a Poissionian distribution of photon arrival times. In this simulation, we have assumed that our detector provides the light-curve data in 1-ms time bins.  The relative time delays are then calculated for each satellite combination. Figure \ref{fig:lc_sample} show examples for output of this light curve simulation, and we can see that the differences in the photon arrival times are well reproduced. The time delays for all combinations of the satellites are determined by the cross-correlation analysis after subtraction of the background. We performed this simulation 100 times, including the Poisson fluctuations in order to estimate the statistical error of the simulation.

\subsection{Localization algorithm}
\begin{figure}[htbp]
\begin{center}
 \rotatebox{-0}{\resizebox{10cm}{!}{\includegraphics{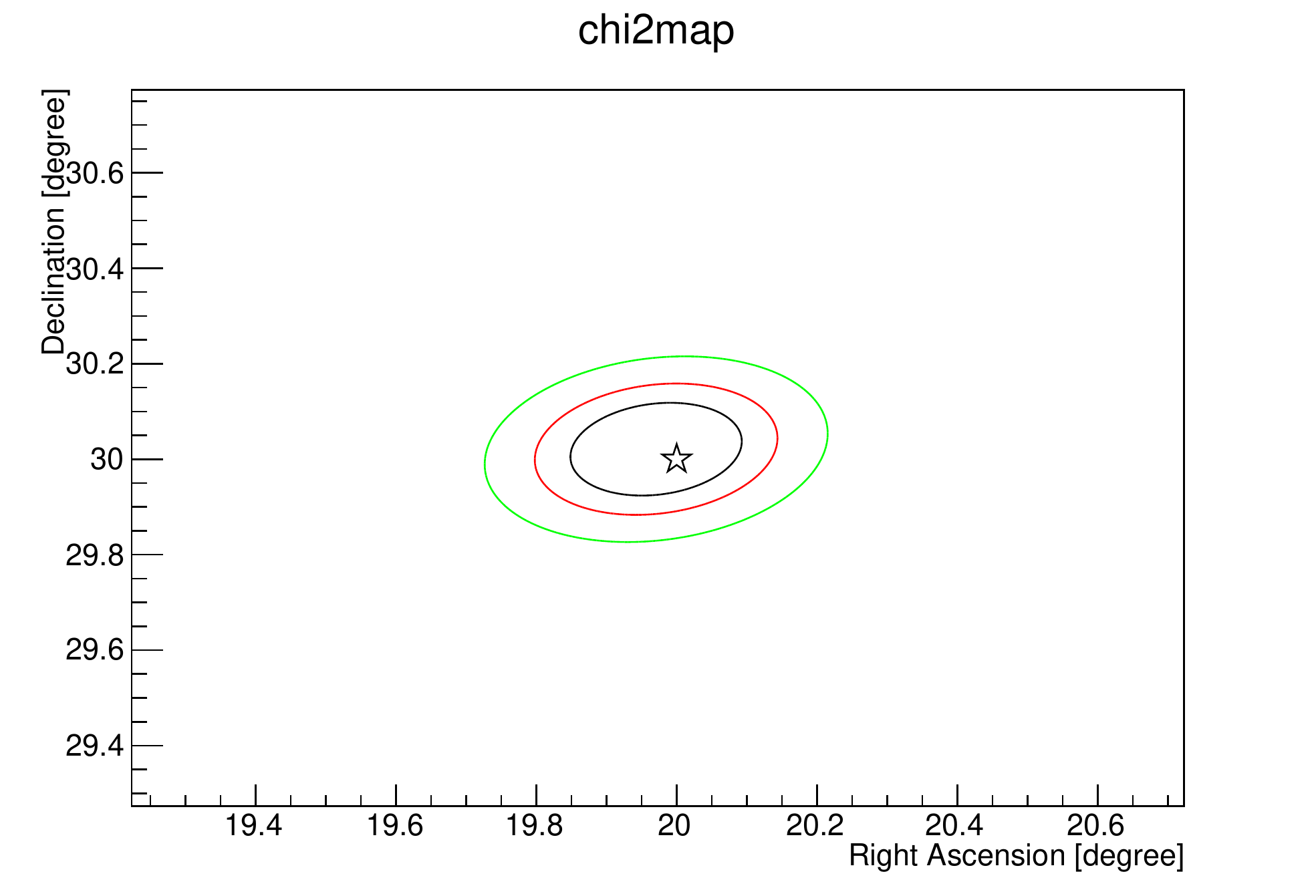}}}
\caption{An example of a simulated localization of the bright SGRB GRB100101988. The black, red, and green contours represent the 1-, 2-, and 3-$\sigma$ confidence regions of our localization. The star marker shows the input position.}
\label{fig:loc_chi2}
\end{center}
\end{figure}

We obtain a set of simulated $\delta t$ values with their errors for all combinations of the visible satellites. We calculate a set of annuli using the triangulation formula, cos$\theta$ = c$\delta t/D$. For the traditional triangulation method utilized by the Interplanetary Network (IPN), a limited set of satellites (typically 3) and a relatively large error for the $\delta t$  enable the intersection area for all the annuli to be determined by eyes. For our case, however, since the number of the satellites and the combinations is much larger than that for the IPN case (typically 15$-$21 combinations by 6$-$7 satellites) and since the error in  $\delta t$ is much smaller (typically,  tens to hundreds of microseconds), it is no longer possible to find the intersection area for all annuli manually. Thus, we have developed other effective algorithms to find "the most probable position"  for the {\it CAMELOT} mission based on a parameterized fitting method.
In principle, from the triangulation formula, the value of cos$\theta$  can be obtained from $\delta t$, if we assume the error of the distance $D$ to be very small in comparison with the other parameters. We define the quantity $\chi^2$ by the following formula, 
\begin{equation}
\chi^2 \equiv \sum_{i=0}^N \frac{\biggl\{\delta t_{\rm sim,i} - {\rm Norm} \times \rm{cos}\theta_{\rm model,i}  (R.A., Dec.) \times \rm D/c \biggr\}^2}{\sigma_{\rm sim,i}^2}, 
\end{equation}
where, $i$ is the number of the combinations of the visible satellites and $\sigma$ is the error of the simulated  $\delta t$. We introduced the normalization parameter, "Norm" to enable us to take into account any systematic uncertainties. This parameter should ideally be unity, and we have confirmed that it is indeed almost unity for all of our simulations. The most probable GRB position in the celestial coordinates (R.A. and Dec.) can be obtained by minimizing the value of this  quantity $\chi^2$. Since we  have found that the region of parameter space in which  $\chi^2$ achieves its minimum is very limited, this minimization procedure does not converge if the initial parameter values are set as free. We therefore scan the $\chi^2$ values  in 5-degrees steps of a coarse parameter space by fixing the normalization parameter to be unity. After finding the parameters that yield the minimum  value of $\chi^2$ in this coarse parameter space, we perform the final  $\chi^2$ minimization by setting the initial parameters to be the best values obtained in the previous coarse search.  The 1-, 2-, and 3-sigma confidence region for this fit are calculated from the coordinates for which the  $\delta \chi^2$  value reaches 2.30, 4.61, and 9.21, respectively, in the finer parameter space with 0.01 degrees steps within $\pm$ 3 degrees around the best-fit position. Figure \ref{fig:loc_chi2} shows an example of this parameterized $\chi^2$ fitting procedure. In this case, we have assumed that the bright {\it Fermi}-GBM short GRB100101988, with a fluence of 1.87 x 10$^{-6}$ erg cm$^{-2}$ and a duration of 1.98 s, occurred at the celestial coordinates (R.A., Dec.)=(20$^\circ$, 30$^\circ$). Localization with a set of 15  values of  $\delta t$ from 6 visible satellite combinations, which ranges from a few milliseconds to 40 milliseconds, gives the result of (R.A., Dec.) = (19.93$\pm$0.07$^\circ$, 29.96$\pm$0.08$^\circ$; errors represent 1-sigma confidence region), which is consistent with the input position to within the 3-sigma confidence level and the 1-sigma localization accuracy corresponds to a scale of several arcminutes. Therefore, this  parametrized $\chi^2$ fitting procedure works well and our concept of localization using a fleet of nanosatellites can provide a localization acuracy of an order of arcminutes for a bright burst.

\subsection{Localization performance}

\begin{figure}[htbp]
\begin{center}
\begin{minipage}{8cm}
 \rotatebox{-0}{\resizebox{6cm}{!}{\includegraphics{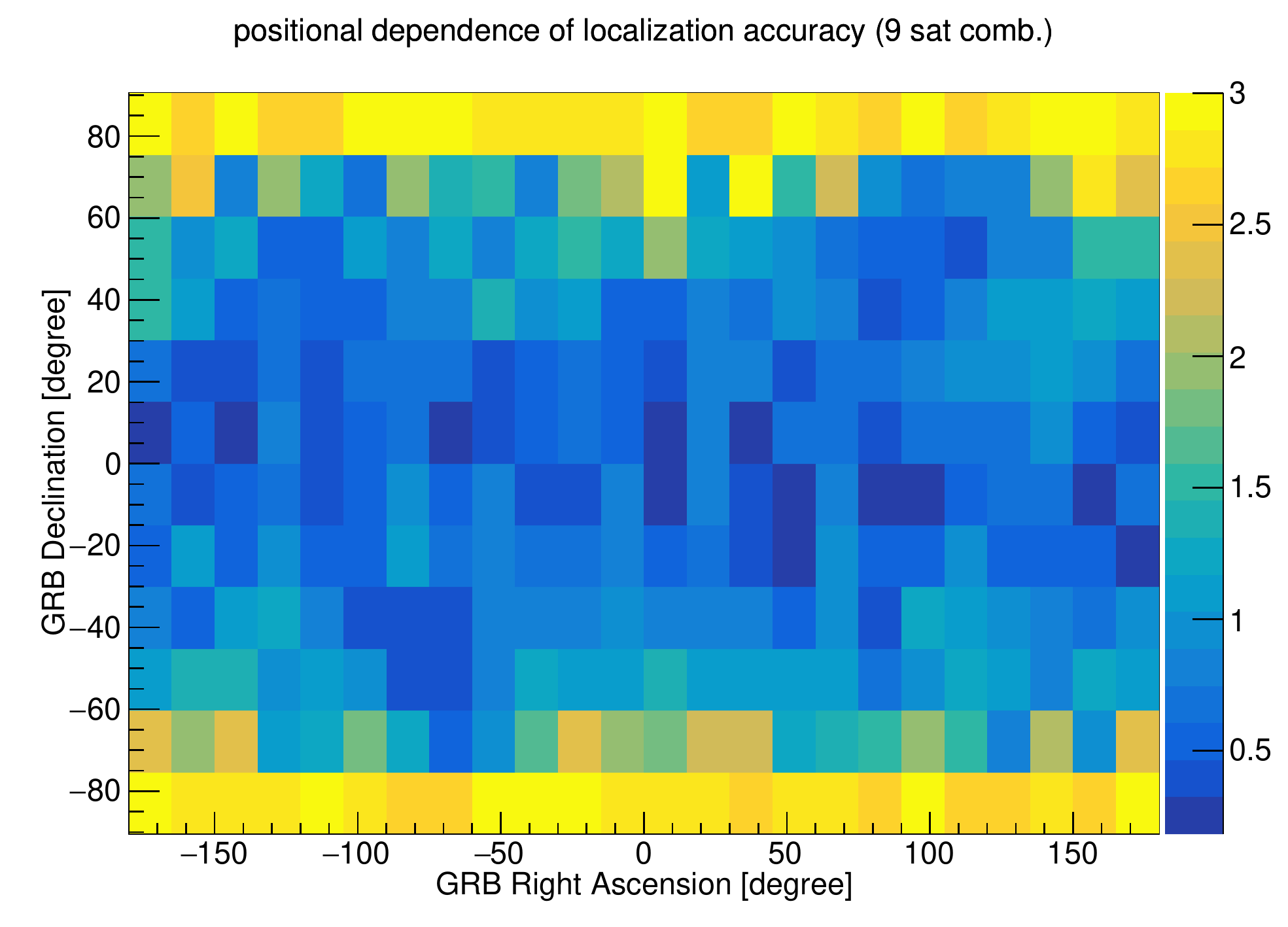}}}
\end{minipage}
\begin{minipage}{8cm}
 \rotatebox{-0}{\resizebox{6cm}{!}{\includegraphics{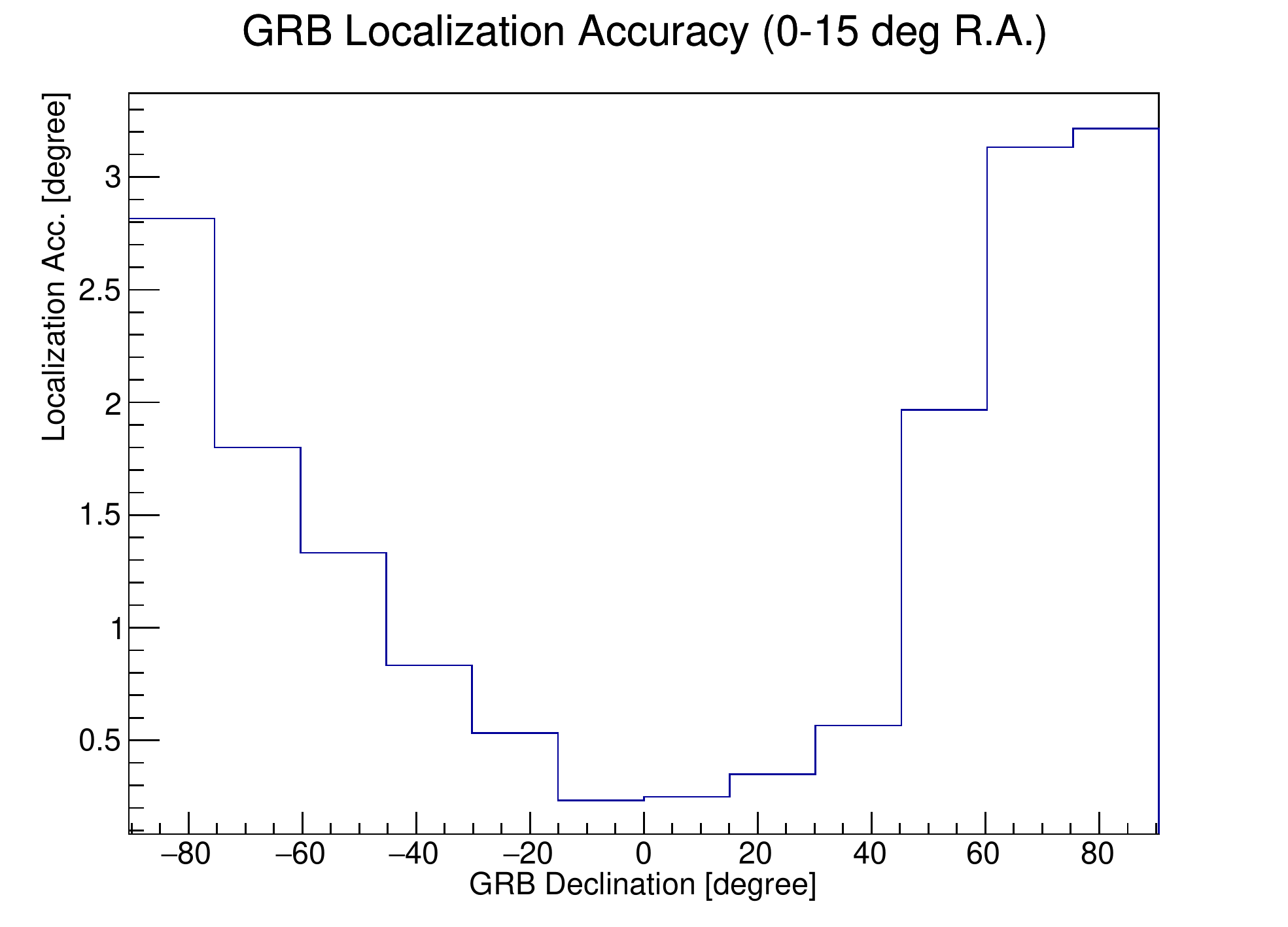}}}
\end{minipage}
\caption{(Left) The GRB incident-position dependence of the localization accuracy in a celestial coordinates for GRB 100929916, with an intermediate fluence of  7.6$\times$10$^{-7}$ erg cm$^{-2}$. Note that the localization accuracy around the polar region is worse since current localization accuracy is determined as a cosine function in the spherical coordinates. The color scale of this figure represents the localization accuracy in degrees. (Right) The projection of this distribution onto the declination axis in the between 0$-$15 degrees of right ascension.}
\label{fig:loc_posdep}
\end{center}
\end{figure}

\begin{figure}[htbp]
\begin{center}
 \rotatebox{-0}{\resizebox{12cm}{!}{\includegraphics{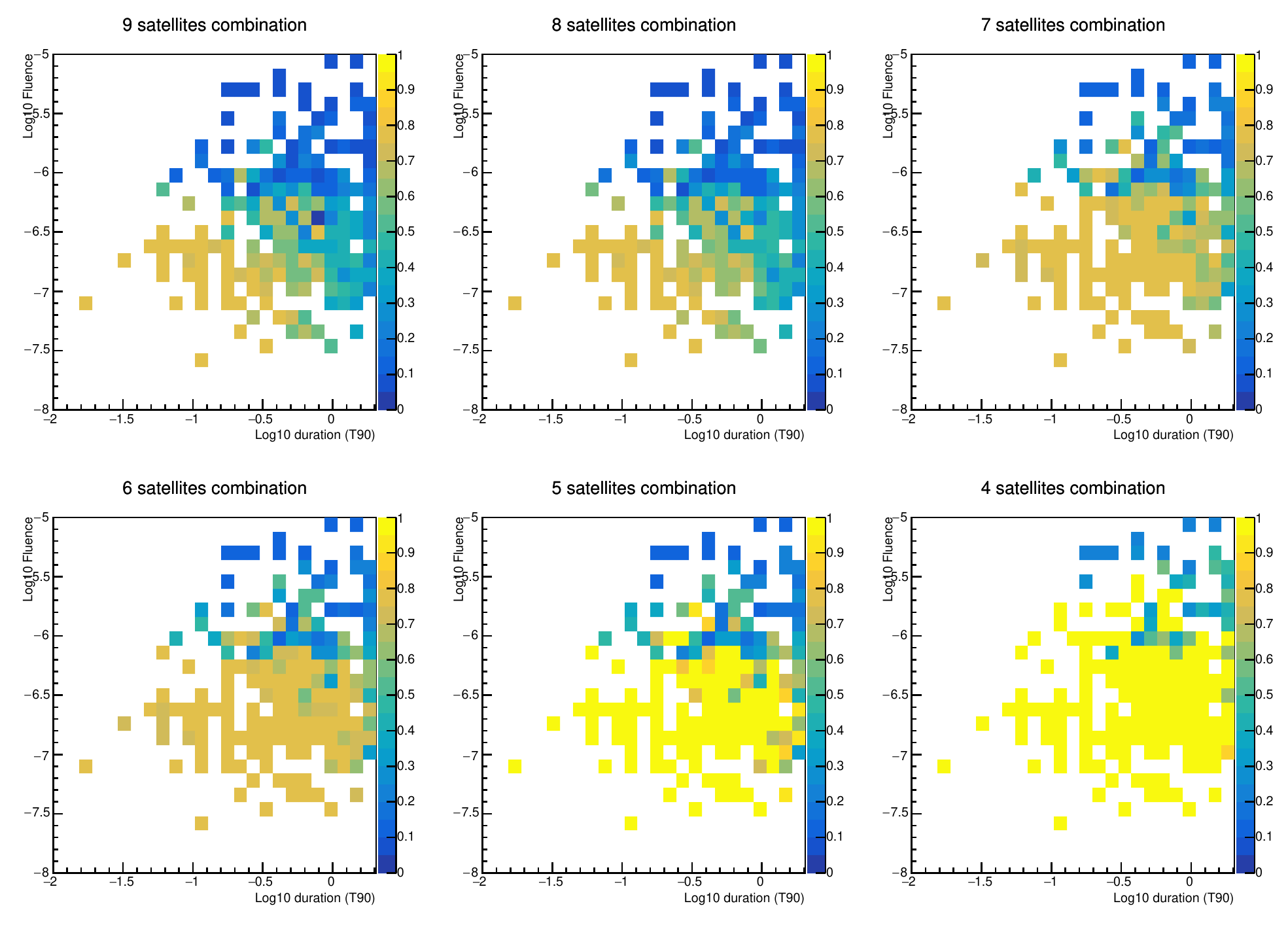}}}
\caption{The  result of our localization analysis. Each figure shows how accurately our {\it CAMELOT} mission localizes GRBs, depending on the GRB duration (horizontal axis) and the GRB brightness (vertical axis). Each panel shows the results for detections by nine to four satellites, respectively. The color scale of these figures represents the localization accuracy in degrees.}
\label{fig:loc_systematic}
\end{center}
\end{figure}

After confirming that our localization algorithm works well, we next discuss the systematic study of the GRB characteristics, which is the most important  for  future GRB observations. In this section, we examine the dependence of the localization accuracy on the GRB brightness and duration. In addition, we have checked the dependence on the number of visible satellites.
As shown in the previous sections, we have used the chi-square-minimization method for the localization, and we have used the fixed GRB position of (R.A., Dec.) = (20$^\circ$, 20$^\circ$) and each satellite position, which means the fixed $\delta t$ value and the effective area for a given satellite, for all the systematic analyses in this section. We have confirmed  that the localization accuracy exhibits no clear dependence on the right ascension direction  and that it is symmetrical around the declination zero degrees, as shown in Figure \ref{fig:loc_posdep}. In addition, by considering the sky coverage, it is possible to regard the declination of 20$^\circ$-40$^\circ$ as a typical direction of the GRB incident angle. For the GRB characteristics used for this study, we have examined all 363 SGRBs (T90 $<$ 2.0 s) listed in the 3rd {\it Fermi}-GBM catalog paper\cite{0067-0049-223-2-28}.  Figure \ref{fig:loc_systematic} displays a summary of this systematics study. It clearly shows that longer and brighter GRBs provide better localization accuracy, as expected from the statistics. Since there is no other clear trend in this figure, the statistical limit is much more important than the shape of the light curve. Another clear trend is that smaller number of satellite combinations make the localization accuracy worse.  For fewer than five visible satellites, the localization accuracy for many SGRBs can be worse than 1 degree. Table \ref{tab:localization_accuracy} shows a summary of localization accuracy as a function of the number of visible satellites. For example, we found that about 13 \% of the {\it Fermi}-GBM SGRBs can be localized to within 20 arcminutes accuracy for a five-satellites detection. By combining this expected localization capability for all available visible satellite patterns as shown in Table~\ref{tab:localization_accuracy}, we find that about 10 SGRBs per year can be localized to within 20 arcminutes accuracy, based on the statistics reported in the 3rd {\it Fermi} catalog.

\begin{table}[h]
\centering
\begin{tabular}{ccccc}
\hline
\hline
Number of  & Visibility       & Fraction $(<10^\prime)$ & Fraction $(<15^\prime)$ & Fraction $(<20^\prime)$ \\
satellites & probability [\%] & [\%] & [\%]                    & [\%]                    \\ [0.5ex]
\hline
9                &  1.8  &  27  &  30  & 37  \\
8                &  7.6  &  26  &  29  & 33  \\
7                &   16  &   5  &  14  & 19  \\
6                &   26  &   2  &   8  & 13  \\
5                &   25  &   1  &   3  & 8   \\
4                &   15  &   1  &   1  & 1   \\
\hline
\end{tabular}
\caption{The expected visibility probabilities of the given number of satellites and the fractions of the number of {\em Fermi}-GBM SGRBs with given localization accuracy. }
\label{tab:localization_accuracy}
\end{table}

\section{CONCLUSIONS}

We propose a fleet of nanosatellites, called the {\it CAMELOT}, for localizing GRBs some of which will be the electromagnetic counterparts of gravitational-wave sources. Considering the anticipated field of view of future
optical telescopes, our mission aims to localize GRBs to within one-degree accuracy with an all-sky coverage. 
Further localization accuracy is of course useful for constraining the number of host-galaxy 
candidates within the field of view. Our new concept, which combines good localization with all-sky coverage
 in the gamma-ray energy band, can be realized by measuring the differences in arrival times of photons
for each satellite with excellent GPS timing synchronization. 
Our current detector design for each satellite utilizes a large, plate-shaped CsI scintillator with an effective area comparable to {\it Fermi}-GBM, and we find that our detector can achieve a good low-energy threshold of 10 keV. Including these results from our detector-evaluation tests, we have developed a framework for determining the localization accuracy for our mission concept. A systematic analysis of our localization accuracy using the 3rd {\it Fermi}-GBM GRB catalog demonstrates that our mission concept can obtain a localization accuracy of tens of arc minutes scale for bright GRBs and that about 10 SGRBs per year can be 
localized to  within 20 arcminutes with our {\it CAMELOT} mission. 

\acknowledgments 
This work was supported by the by Hiroshima university, JSPS KAKENHI Grant Number 17H06362, the Lend\"uet LP2016-11 and LP2012-31 grants awarded by the Hungarian Academy of Sciences, as well as by GINOP-2.3.2-15-2016-00033.

\bibliography{mybibfile_v20180524} 
\bibliographystyle{spiebib} 

\end{document}